\documentclass[prf,superscriptaddress,amsmath,amssymb,nofootinbib,floatfix,notitlepage,longbibliography]{revtex4-1}

\usepackage{amsfonts, amsmath, amssymb, array, eucal, mathtools, xfrac,
booktabs, array, color, natbib, graphicx, siunitx, tikz, pgfplots, subcaption,graphicx}

\pgfplotsset{compat=1.8, every axis/.append style={font=\footnotesize}}
\pgfplotsset{
        compat=1.8,
        tick scale labels in axis labels/.code={
            \pgfkeysgetvalue{/pgfplots/xtick scale label code/.@cmd}\temp
            \pgfkeyslet{/pgfplots/xtick scale label code orig/.@cmd}\temp
            \pgfkeysalso{
                xtick scale label code/.code={
                    \gdef\xTickScale{##1}
                },
                every axis/.append style={
                    tick scale binop={},
                    xlabel/.add = {}{
                        (\pgfplotsset{xtick scale label code orig=\xTickScale}\,m)
                    },
                },
            }
        },
    }

    \pgfplotsset{
        compat=1.8,
        tick scale labels in axis labels/.code={
            \pgfkeysgetvalue{/pgfplots/ytick scale label code/.@cmd}\temp
            \pgfkeyslet{/pgfplots/ytick scale label code orig/.@cmd}\temp
            \pgfkeysalso{
                ytick scale label code/.code={
                    \gdef\yTickScale{##1}
                },
                every axis/.append style={
                    tick scale binop={},
                    ylabel/.add = {}{
                        $\cdot$\pgfplotsset{ytick scale label code orig=\yTickScale}
                    },
                },
            }
        },
    }
\usetikzlibrary{shapes,patterns,arrows,arrows.meta,fadings,calc,positioning,mindmap,trees,shadows}
\tikzset{
    double arrow/.style args={#1 colored by #2 and #3}{
    -stealth,line width=#1,#2, 
    postaction={draw,-stealth,#3,line width=(#1)/3,
                shorten <=(#1)/3,shorten >=2*(#1)/3}, 
  }
}

\definecolor{midori}{rgb}{0.0, 0.6, 0.0}

\begin{document}
\title{Microtransformers: controlled microscale navigation with flexible robots}
\author{Thomas D. Montenegro-Johnson}
\affiliation{School of Mathematics, University of Birmingham, Edgbaston,
Birmingham, UK, B15 2TT}
\date{\today}

\begin{abstract}
Artificial microswimmers are a new technology with promising microfluidics and biomedical applications, such as directed cargo transport, microscale assembly, and targeted drug delivery. A fundamental barrier to realising this potential is the ability to independently control the trajectories of multiple individuals within a large group. A promising navigation mechanism for ``fuel-based'' microswimmers, for example autophoretic Janus particles, entails modulating the local environment to guide the swimmer, for instance by etching grooves in microchannels. However, such techniques are currently limited to bulk guidance. This paper will argue that by manufacturing microswimmers from phoretic filaments of flexible shape-memory polymer, elastic transformations can modulate swimming behaviour, allowing precision navigation of selected individuals within a group through complex environments.
\end{abstract}

\maketitle

\section{Introduction}

Driven by advances in manufacturing techniques~\citep{walther2008janus}, the theory of biological locomotion~\citep{lauga2009hydrodynamics,katuri2016artificial}, and myriad biomedical~\citep{nelson2010microrobots} and microfluidics applications~\citep{maggi2016self}, there has been a recent surge of interest in artificial propulsion mechanisms at microscopic scales. Often, these propulsion mechanisms are inspired by the natural world; \citet{dreyfus2005microscopic} created a sperm-like swimmer by attaching a flexible magnetic filament to a red blood cell, \citet{Zhang2009a,Zhang2009b} created a bacterium inspired microswimmer from a microscale helix with a magnetic head, and \citet{tierno2008controlled} exploited boundary screening effects in a similar manner to vertebrate nodal cilia~\citep{smith2008fluid} to create a microswimmer comprising two linked beads of different sizes floating above a flat plate. 

In each of these examples, propulsion was achieved via the application of an external magnetic field, and as such these may be referred to as \textit{externally actuated} microswimmers. Microswimmers propelled by oscillating bubbles driven by applied ultrasound~\citep{bertin2015propulsion} also fall into this category. In contrast, \textit{fuel-based} microswimmers utilise fuel in their immediate environment in order to self-generate propulsion. For instance, \citet{williams2014self} were able to selectively attach cardiomyocytes to a polymer filament to create a self-propelled ``biohybrid'' swimmer.




Whilst externally-actuated microswimmers give the user direct control over their trajectory, the control of individuals within a swarm is limited, requiring the development of complex strategies~\citep{katsamba2016micro}; { direct one magnetic swimmer to the left, and others in the group will also change direction (Fig.~\ref{fig:cartoon}a). Furthermore,} ungainly and expensive equipment is usually required to drive them. In contrast, fuel-based microswimmers do not require such equipment, but their trajectory is not user-defined; it arises from complex physical interactions between the swimmer, fluid, and boundary features. The absence of precision control for fuel-based microswimmers is a major drawback for many proposed applications, such as microscale cargo transport. This study will propose a new mechanism of precision trajectory control for fuel-based microswimmers driven by phoretic effects.

Phoretic microswimmers generate propulsion via gradients of a field, for instance temperature~\citep{jiang2010,bickel2013}, electrical charge~\citep{nourhani2015self}, or chemical solute concentration~\citep{golestanian2005}. In autophoresis, the microswimmer self-generates solute concentration gradients via differential surface reaction, usually achieved through patterning of a catalyst~\citep{paxton2004}. The canonical microswimmer is the ``Janus'' particle~\citep{walther2008janus}: an inert sphere or rod half-coated in a catalyst for the solute, for instance platinum in hydrogen peroxide~\citep{howse2007,Ebbens2011}.

Since the trajectory of autophoretic particles emerges from complex interactions between solute and fluid dynamics, it can be affected by both particle shape, ambient flows, and domain boundaries. As such, autophoretic systems naturally lend themselves to ``physics-based'' control. Approaches for physics-based control of autophoretic microswimmers have hitherto focused on modulating the environment to guide the swimmer. For instance, Janus particles can be guided along grooves in channels~\citep{simmchen2016topographical}, or across applied flows~\citep{katuri2017cross}. However, this study will focus on modulating the swimmer to navigate the environment; a flexible phoretic filament, fabricated from shape-memory polymers, could achieve precision navigation through a motion analogous to bacterial run-and-tumble, transforming between pumping, translating, and rotating modes of operation {(Fig.~\ref{fig:cartoon}b)}. The approach herein is inspired both by nature, and recent experimental work demonstrating the dynamic self-assembly of rigid ``Saturn'' rods into translating and rotating multiparticle structures~\citep{wykes2016dynamic}, theoretical work on eloganted Janus particles~\citep{michelin2017geometric}, and work on flexible chains of linked Janus particles undergoing self-induced oscillations~\citep{vutukuri2017rational}.

\section{Modelling}

\begin{figure}
    \begin{center}
    	\includegraphics{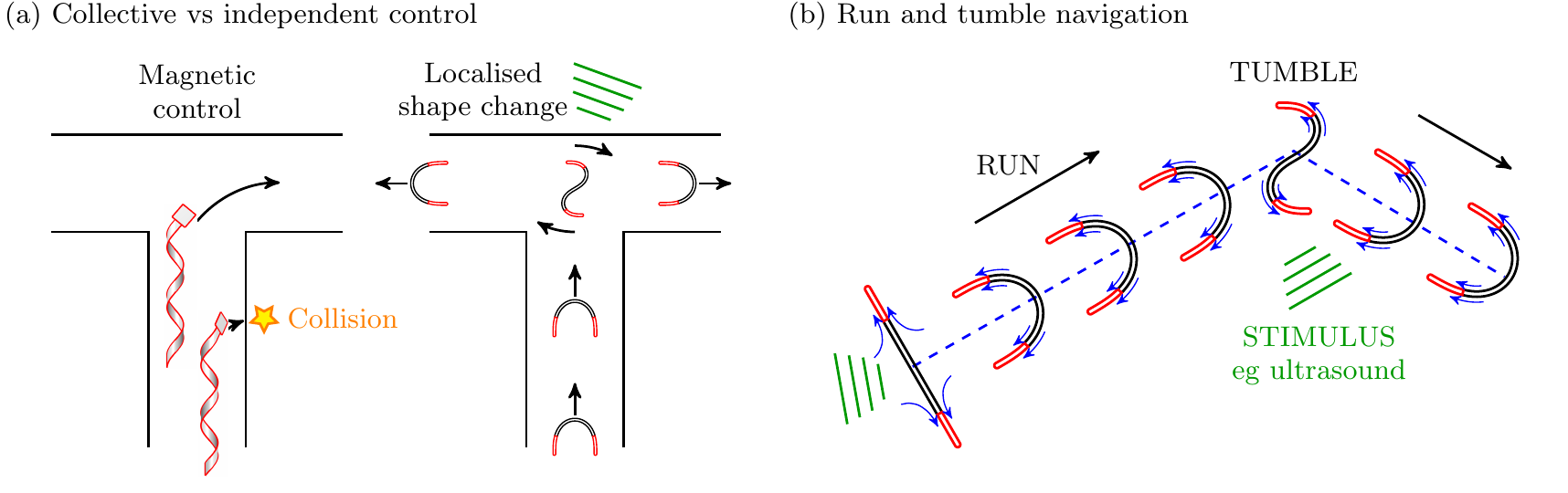}
    \end{center}
    \caption{ a) Many applications require independent control of microswimmers in a group. Left: Directing the leading helical microswimmer~\citep{Zhang2009a} to the right by varying the applied magnetic field causes the trailing swimmer to crash into the channel wall. Right: By selectively heating filaments at the desired location, microtransformers in a swarm could be independently directed to different targets. b) Changing the shape of a microtransformers switches between pumping, translating, and rotating modes, allowing navigation via a series of runs, punctuated by on-the-spot reorientations.}
    \label{fig:cartoon}
\end{figure}

The mechanism for autophoretic propulsion of a filament is as follows. The chemical reaction of a surrounding solute is catalysed on the filament surface, yielding local product concentration gradients; this property is the filament's \emph{activity}. This differential interaction drives local pressure imbalances in a thin layer around the filament, creating a surface slip flow; this property is the filament's \emph{mobility}.

Thus, a straight filament with both ends coated in catalyst should create locally lower concentrations of solute at either end. This gradient of high to low concentration running from the middle to either end creates a surface slip flow that pulls-in solute equatorially, and pushes it out either pole (or vice-versa, depending on the sign of the filament's mobility). In this way, the straight filament acts as a pump. 

By fixing the filament into the 1\textsuperscript{st}-mode U-shape, the product concentration gradient in the surface is little affected, and so flow is now directed out of the top of the U, allowing the filament to translate (Fig.~\ref{fig:cartoon}). Finally, by fixing the filament into the 2\textsuperscript{nd}-mode S-shape, flow is directed out of either pole creating a torque which is balanced by rotational velocity. By fabricating this filament from, {for instance, reversible} shape-memory polymer {\citep{behl2013reversible}} and programming the U- and S-shapes~{\citep{behl2007shape}}, the filament could be stimulated to switch between rotational and translational motion (Fig.~\ref{fig:cartoon}b) { via targeted heating, for instance with an infrared laser~\citep{leng2011shape} or} focused ultrasound~\citep{bhargava2017focused}, to navigate along a desired trajectory. Targeting specific swimmers in a group could thus allow independent control of individuals within a swarm to perform complex tasks in concert (Fig.~\ref{fig:cartoon}a). We will now test this hypothesis numerically.

\subsection{Autophoretic motion}

We consider neutral solute self-diffusiophoresis, in the absence of electrokinetic effects~\citep{ebbens2014,brown2014}, in the zero P\'eclet number continuum framework of \citet{golestanian2005}. In this framework, solute advection is neglected and the solute concentration $c$ is decoupled from the flow dynamics, yielding Laplace's equation,
\begin{equation}\label{eq:laplace}
D\nabla^2 c=0.
\end{equation}
At the filament's surface, solute is released or absorbed with constant flux,
\begin{equation}
    D\mathbf{n}\cdot\boldsymbol{\nabla}{c} = -\mathcal{A}(\mathbf{x}).
\end{equation}
for filament activity $\mathcal{A}$, {outward normal $\mathbf{n}$,} and solute diffusivity $D$. The phoretic slip velocity boundary condition for the flow problem is then given by the surface gradient of the solution to the Laplace problem,
\begin{equation}\label{eq:mobility}
    \mathbf{u}_s = \mathcal{M}(\mathbf{x})(\mathbf{I} - \mathbf{n}\mathbf{n})\cdot
    \boldsymbol{\nabla} c \quad \mbox{on } S,
\end{equation}
for filament mobility $\mathcal{M}$ {and surface $S$. Since this propulsive slip flow is always tangential to the filament surface, it is thus possible to redirect the propulsion, and thereby change the swimming behaviour, simply by deforming the filament (ie altering $S$).} At microscopic scales, { the resultant fluid flow field $\mathbf{u}$ can then} be calculated via the dimensionless Stokes flow equations,
\begin{equation}
    \nabla^2\mathbf{u} = \boldsymbol{\nabla}p, \quad
    \boldsymbol{\nabla}\cdot\mathbf{u} = 0,
    \label{eq:stokes_flow}
\end{equation}
{ where $p$ is the fluid pressure.} The problem is non-dimensionalised using the filament length $L$, $\mathcal{A}L/{D}$ and $\mathcal{AM}/{D}$ as characteristic length, concentration and velocity scales, respectively.

\subsection{Elastostatic shapes}
In order to model shapes that may be manufactured simply by buckling { and subsequent fixation of a shape memory filament} between two tweezers, we solve the Euler-Bernoulli equation~\citep{howell2009applied} for the configuration of a geometrically nonlinear, inextensible beam,
\begin{equation}
	EI\frac{\mathrm{d}^2\theta}{\mathrm{d}s^2} + N_0\cos\theta - T_0\sin\theta = 0,
\end{equation}
where $\theta$ is the tangent angle of the beam, $s$ is the arclength along the centreline. The normal and tensile forces $N_0,T_0$ respectively are applied at $s=1$, with equal and opposite forces applied at $s=0$, and the constant $EI$ is known as the bending stiffness of the beam. 

Solving the beam equation subject to clamped boundary conditions at the proximal end
\begin{equation}
	\theta(0) = 0, \quad \left.\frac{\mathrm{d}\theta}{\mathrm{d}s}\right|_{s=0}\!\!\!\!\!\!\!\! = 0,
\end{equation}
we adjust the applied normal and tensile forces with a shooting method in order to staisfy 
\begin{equation}
	\theta(1) = -\pi,0, \quad \left.\frac{\mathrm{d}\theta}{\mathrm{d}s}\right|_{s=0}\!\!\!\!\!\!\!\! = 0,
\end{equation}
at the distal end, for the first and second buckling modes respectively. The calculated shapes shown in figure~\ref{fig:cartoon} are then used to solve the steady-state autophoretic motion of the microtransformer.

\subsection{Numerical solution of the diffusiophoresis problem}

The equations governing the filament's autophoretic motion will be solved using a regularised boundary element method~\citep{montenegro2015regularised}, which we will now summarise. For the dimensionless Laplace problem, { the concentration at a point $\mathbf{x}_0$ is given by} the boundary integral,
\begin{equation}
    \lambda c(\mathbf{x}_0) =
    \int_{S}c(\mathbf{x}) \mathbf{K}^\epsilon(\mathbf{x},\mathbf{x}_0) 
\cdot \mathbf{n}(\mathbf{x})
  - \frac{\partial c(\mathbf{x})}{\partial\mathbf{n}}G^\epsilon (\mathbf{x},\mathbf{x}_0)\,\mathrm{d}S_{x},
    \label{eq:diff_bem}
\end{equation}
{ where the} regularised sink $G^\epsilon$ and source dipole $\mathbf{K}^\epsilon$ { located at $\mathbf{x}$ are given by,}
\begin{equation}\label{eq:reg_gf}
    G^\epsilon(\mathbf{x},\mathbf{x}_0) = -\frac{2r^2 + 3\epsilon^2}{8\pi
    r_\epsilon^3}, \quad 
    K^\epsilon_j(\mathbf{x},\mathbf{x}_0) = r_j\frac{2r^2 +
    5\epsilon^2}{8\pi r_\epsilon^5},
\end{equation}
with $r_j = (\mathbf{x} - \mathbf{x}_0)_j,\ r = |\mathbf{x} - \mathbf{x}_0|$. {These regularised Green's functions correspond to a specific} regularisation of the Dirac $\delta$-function given by~\citep{Cortez05},
\begin{equation}
    \phi_\epsilon(\mathbf{x} -
    \mathbf{x}_0) = \frac{15\epsilon^4}{8\pi r_\epsilon^7}, \quad r_\epsilon^2 =
    r^2 + \epsilon^2.
    \label{eq:blob_choice}
\end{equation}
Provided the regularisation $\epsilon$ is much smaller than $\kappa$, the mean local curvature of the surface \citep{montenegro2015regularised}, the constant $\lambda\approx 0,1/2,1,$ when the evaluation point $\mathbf{x}_0$ is inside, on, or outside the boundary respectively.

While it is possible~\citep{montenegro2015regularised} to calculate the surface concentration gradient by differentiating equation~\eqref{eq:diff_bem}, the geometric regularity of the filament mesh allows us to fit a cubic spline interpolant using the matlab function \texttt{csape}, with periodic boundary conditions in the azimuthal direction, to the calculated concentration values, in a similar manner to previous work in two-dimensions~\citep{michelin2015geometric}. This method gives a fast and accurate rendering of the surface concentration gradient, { which is used to provide the slip velocity boundary condition for the Stokes flow problem.}


{ To solve the Stokes flow problem, we employ the boundary integral,}
\begin{equation}
    \lambda u_j(\mathbf{x}_0) = \int_S
     S_{ij}^\epsilon(\mathbf{x},\mathbf{x}_0)f_i(\mathbf{x}) -u_i(\mathbf{x})T_{ijk}^\epsilon(\mathbf{x},\mathbf{x}_0)
    n_k(\mathbf{x})\,\mathrm{d}S_x,
    \label{eq:reg_bem}
\end{equation}
of regularised stokeslets $S_{ij}^\epsilon$ and their associated stress fields $T_{ijk}^\epsilon$~\citep{Cortez05}
\begin{equation}
    S_{ij}^\epsilon(\mathbf{x},\mathbf{x}_0) = \frac{\delta_{ij}(r^2 +
    2\epsilon^2) + r_i r_j}{8\pi r_\epsilon^3}, \quad
    T_{ijk}^\epsilon(\mathbf{x},\mathbf{x}_0) = -\frac{6r_i r_j
    r_k}{8\pi r_\epsilon^5} -\frac{3\epsilon^2\left(r_i\delta_{jk} + r_j\delta_{ik} + r_k\delta_{ij}\right)}{8\pi r_\epsilon^5}.
\end{equation}
{Specifying that the velocity on the surface is given by the combination of phoretic slip velocity $\mathbf{u}_s$ and rigid body motion arising from the unknown translational and rotational swimming velocities $\mathbf{U},\boldsymbol{\Omega}$, we make the substitution $\mathbf{u}(\mathbf{x}) = \mathbf{u}_s(\mathbf{x}) +
    \mathbf{U} + \boldsymbol{\Omega}\wedge(\mathbf{x} - \mathbf{x_c})$ for $\mathbf{x_c}$ the filament centre of mass, and we find
\begin{equation}
    \lambda u^s_j(\mathbf{x}_0) + \int_S u^s_i(\mathbf{x})T_{ijk}^\epsilon(\mathbf{x},\mathbf{x}_0)
    n_k(\mathbf{x})\,\mathrm{d}S_x = \int_S S_{ij}^\epsilon(\mathbf{x},\mathbf{x}_0)f_i(\mathbf{x}) -[U_i + (\boldsymbol{\Omega}\wedge[\mathbf{x}_0 - \mathbf{x}_c])_i]T_{ijk}^\epsilon(\mathbf{x},\mathbf{x}_0)
    n_k(\mathbf{x})\,\mathrm{d}S_x,
    \label{eq:swim_bem}
\end{equation}
for unknown surface tractions $\mathbf{f}$. The} unknown translational and rotational velocities are found by enforcing the constraints that zero net force or torque act upon the filament,
conditions
{
\begin{equation}
    \int_S \mathbf{f}(\mathbf{x})\,\mathrm{d}S_x = \mathbf{0},
    \quad
    \int_S (\mathbf{x} - \mathbf{x}_c)\wedge\mathbf{f}(\mathbf{x})\,\mathrm{d}S_x = 
    \mathbf{0}.
\end{equation}}
These boundary integral equations are discretised over a surface mesh of piecewise quadratic triangles and solved using the linear panel boundary element method described in the author's previous work~\citep{montenegro2015regularised}, which is available for free download from the Matlab file exchange\footnote{http://uk.mathworks.com/matlabcentral/ profile/authors/5102158-thomas-montenegro-johnson}.

\section{Results}

\begin{figure*}
     \begin{center}
     	\includegraphics{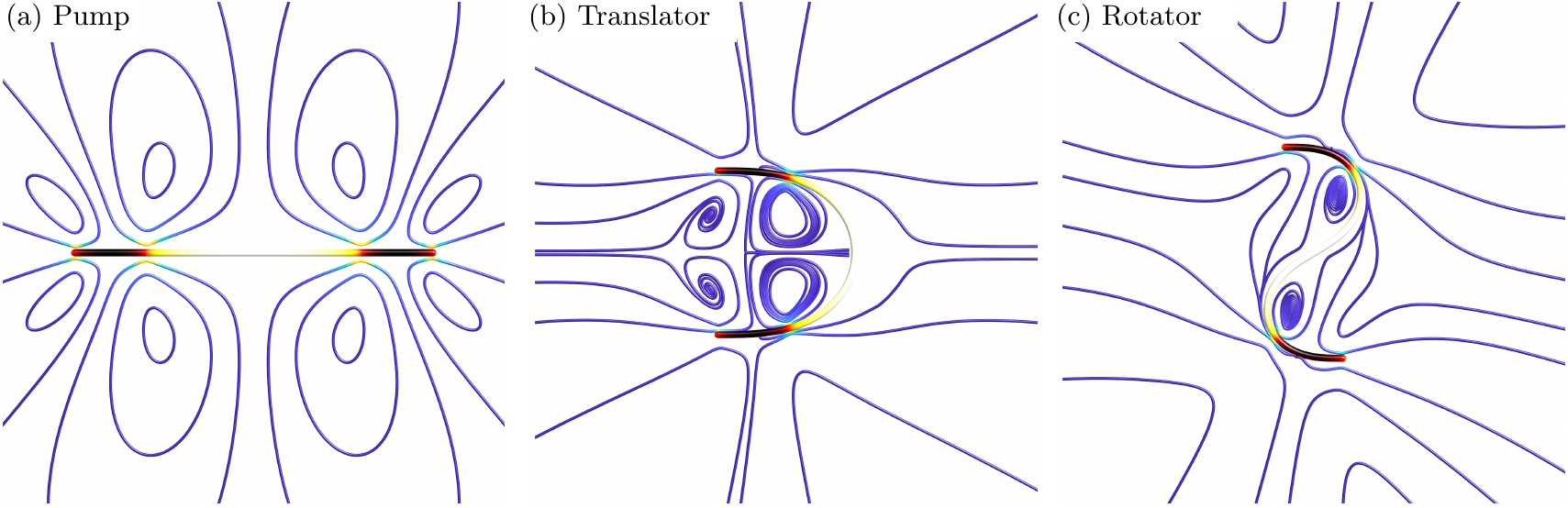}
     \end{center}
    \caption{Schematic surface solute concentration and flow streamlines of a flexible phoretic filament with catalytic endcaps, semi-coverage $s_c = 0.2$, and uniform positive mobility, in (a) the pumping configuration, (b) the translating configuration, and (c) the rotating configuration.}
    \label{fig:streamlines}
\end{figure*}

A typical Janus particle is a polystyrene bead, half-coated in platinum, immersed in hydrogen peroxide. The platinum acts as a catalyst for the reduction of hydrogen peroxide into water and oxygen, so that the concentration of solute is locally lower on near the platinum face. The bead has positive mobility, and so swims with the polystyrene face forwards. We will model a similar filament, with the simplifying assumptions of positive uniform mobility $\mathcal{M}=1$, zero activity on the inert midpiece of the filament, and positive activity on catalytic caps at either end,
\begin{equation}
\mathcal{A} = \begin{cases}
      0, &\text{for}\ \ s_c < s < 1-s_c, \\
      1, & \text{otherwise,}
    \end{cases}
    \label{eq:forward_trans}
\end{equation}
where $s$ is the filament arclength, and $s_c$ is the location of the proximal boundary between the active cap and inert midpiece. By varying $s_c$, the microtransformer can be optimised for pumping, translation, or rotation. In the following simulations, we consider a filament with a slenderness $d/L = 1/50$, for $d$ the filament diameter. The surface concentration and flow streamlines for such a filament are shown for the pumping, translating, and rotating configurations in figure~\ref{fig:streamlines}.

\begin{figure*}[t]
    \begin{center}
		\includegraphics{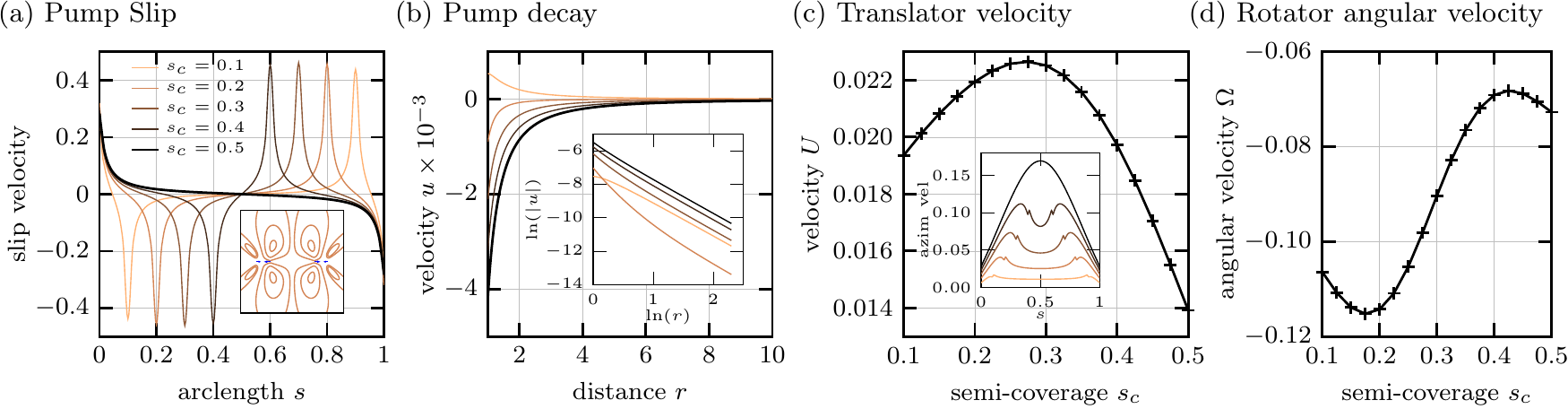}
    \end{center}
    \caption{a) The tangential slip velocity of the pump with catalytic end caps and uniform, positive, mobility, as a function of arclength $s$. The maximum speed occurs at the boundary between the catalytic and inert portions of the filament for all intermediate semi-coverages $s_c$, in an analogous manner to elongated Janus particles~\citep{michelin2017geometric}. Inset shows streamlines from four regularised stokeslets mimicking the flow for $s_c = 0.2$ (Fig.~\ref{fig:streamlines}a). b) Flow decay as a function of distance from the filament centre along the $x$-axis, showing a sign change in the direction of pumping as the cap semi-coverage is increased. The logarithmic plot (inset) demonstrates stresslet-like $1/r^2$ decay. c) The swimming velocity of the translating configuration as a function of cap semi-coverage $s_c$ for catalytic end caps and uniform, positive mobility. Swimming is bend-first. Inset shows the maximum azimuthal velocities as a function of arclength $s$ for $s_c = 0.1$ (lightest), $0.2,0.3,0.4,$ and $0.5$ (darkest). d) The angular velocity of the same filament in the rotating configuration, as a function of $s_c$, showing clockwise motion with ends trailing.}
    \label{fig:pump}
\end{figure*}

Beginning in the straight configuration, we note that our initial na\"{i}ve prediction of the pump's behaviour (Fig.~\ref{fig:cartoon}) does not capture the full behaviour of the system (Fig.~\ref{fig:streamlines}a). In an analagous manner to theoretical calculations of elongated rod-like Janus particles~\citep{michelin2017geometric}, the reduced confinement at the filament poles results in lower product concentration at either end, so that the minimum solute concentration is located at $0<s_1<s_c$ and $1-s_c<s_2<1$. As such, in addition to the surface slip flow from the equator to the poles, there is an opposing flow from the pole inwards towards the equator (Fig.~\ref{fig:pump}a), and the flow field is given by the complex streamline pattern shown in figure~\ref{fig:streamlines}a. Since this inwards flow counteracts the stronger outward flow, the efficacy of the pump will be lower than initially anticipated, and the direction {in which fluid far from the filament is pumped} may flip depending on the value of $s_c$. 

An intuitive way to understand this behaviour is by modelling the filament as four immersed forces, located along the filament centreline at points where the surface slip velocity takes its maximum $s = 0,s_c,1-s_c,1$ (Fig.~\ref{fig:pump}a), from which we can recover a remarkably similar streamline pattern to the full simulation (Fig.~\ref{fig:pump}a, inset). This result suggests that the pump exerts a net stress on the fluid, and that { far from the filament the flow is given by the symmetric part of the force dipole, the stresslet,
	\begin{equation}
		u_i(\mathbf{x}_0) = \frac{3r_ir_jr_k}{r^5}g_{jk},
	\end{equation}
	where the symmetric tensor $g$ contains information about the orientation and strength of the pump. Thus, we see that the flow} decays in the manner $|\mathbf{u}| = A/r^2$, for $r$ the distance to the centre of the pump. 
	
	For a straight filament, centred at the origin and oriented along the $x$-axis, the flow velocity as a function of distance along the $x$-axis can then be used calculate the value of the constant $A$ ({ the strength of the pump}) for changing cap semi-coverage (Fig.~\ref{fig:pump}b), for which we clearly see that the optimal pump is in fact the uniformly active filament with $s_c=0.5$ (Fig.~\ref{fig:pump}b, inset), where for positive mobility fluid is pulled in from either pole and pushed out radially from the equator. {However, for small $s_c$, the pump strength $A$ changes sign, and so flow is pumped in the opposite direction.} Since the pumping strength goes through a sign change, there must exist an intermediate value of $s_c$ for which the straight filament exerts zero stress on the fluid, and the far field decay is at most quadrupolar $|\mathbf{u}| = A/r^3$, which we find to be $s_c\approx 0.18$.

Note that it is possible to design a pump that utilises these end-effects to improve efficiency. By shifting the catalytic portion of the filament to the mid region
\begin{equation}
\mathcal{A} = \begin{cases}
      1, &\text{for}\ \ s_c < s < 1-s_c \\
      0, & \text{otherwise,}
    \end{cases}
\end{equation}
the solute concentration becomes monotonically increasing from the centre to either pole, and the filament is better approximated by two (rather than four) equal and opposite forces, or in the far-field a point stress. However, these pumps do not perform radically better than their counterparts in figure~\ref{fig:pump} and so we conclude that these end effects are not paramount when designing flexible phoretic filaments with uniform mobility. 

We might reasonably expect that the strongest pump corresponds to the fastest translator, however, as with tori~\citep{schmieding2017autophoretic}, this is not the case. The swimming speed of the microtransformer in the translating configuration is shown as a function of the cap semi-coverage $s_c$ in figure~\ref{fig:pump}c. For catalytic end-caps and uniform positive mobility across the filament, the microtransformer swims with the bend first with its ends trailing behind. The fastest translator occurs at approximately $s_c = 0.275$ shown in figure~\ref{fig:pump}c. Unlike the pump, the translator does not undergo a switch in sign, and so always travels in the same direction.

There is also now an important curvature/confinement effect; inside the U-bend, fresh solute has less freedom to diffuse due to confinement, resulting in locally lower concentration than at the same arclength on the outside of the bend. The result is an azimuthal slip flow from positive to negative curvature that aids in propulsion (Fig.~\ref{fig:pump}c, inset), that is comparable in speed to the tangential slip over much of the filament (Fig.~\ref{fig:pump}a). In light of the pump results, we can quantify the importance of this azimuthal flow by once again considering 
\begin{equation}
\mathcal{A} = \begin{cases}
      1, &\text{for}\ \ s_c < s < 1-s_c \\
      0, & \text{otherwise,}
    \end{cases}
    \label{eq:reversed_trans}
\end{equation}
with uniform negative mobility. In this case, the tangential slip flow remains comparable for intermediate semi-coverage, though without contrary flow at the ends, but the azimuthal flow is reversed. The filament again swims bend-first, and the maximum again occurs at approximately $s_c = 0.275$, however the swimming velocity is just two-fifths that of the previous case because of this reversed flow. { This result can in fact be deduced via the linearity of the governing equations; the swimming velocity of the reversed configuration~\eqref{eq:reversed_trans} for given $s_c$ is equal to the velocity of configuration~\eqref{eq:forward_trans} with the same $s_c$, minus the velocity of the fully catalytic swimmer.} While it is expected that effect of this azimuthal flow will reduce as the filament becomes more slender, given that the ratio of length to diameter of filaments in these results is $1/50$ it seems reasonable that any slender body theory of phoretic filaments to be developed should take this curvature flow effect into consideration.

Finally, we see that the filament in the rotating configuration behaves as expected from the translator results; it rotates clockwise with each end trailing behind (Fig.~\ref{fig:pump}d). This rotation is again strongly aided by azimuthal flow, and the optimal semi-coverage is $s_c \approx 0.175$. The choice of cap semi-coverage should therefore be tailored to the desired application: for cargo transport~\citep{palacci2013photoactivated} optimal translation may be preferable, whereas for mobile mixing~\citep{ceylan2017mobile} optimal rotation may be preferable. { However, we note that for all cap semi-coverages translating and rotating behaviours are observed, suggesting that this mode-switching navigation strategy should be reasonably robust to small variations or errors in chemical patterning that might arise during manufacture.} 
\section{Discussion}

In this work, we have demonstrated that a flexible filament with fixed chemical patterning can change its autophoretic behaviour by changing its geometric configuration. When straight, the filament can act as a pump, when bent into a U-shape, a translator, and when bent into an S-shape, a rotator. By constructing the filament from shape-memory polymer, translating and rotating shapes could be pre-programmed, so that a user could actively switch between modes, perhaps via local heating from focused ultrasound~\citep{bhargava2017focused} or by laser. This switching could enable precision navigation through complex microenvironments of individuals within a swarm via a ``run-and-tumble'' series of straight runs followed by on-the-spot reorientations. While applied here to autophoretic propulsion, this method of control via selective shape transformation is in fact broadly applicable to other means of propulsion, which may be preferable for biomedical applications. For instance, if the catalytic end-caps were to be replaced by encapsulated bubbles driven by ultrasound~\citep{bertin2015propulsion}, this swimmer would drive very similar flows to the phoretic swimmer in all three configurations, with the additional benefit of biocompatibility~\citep{ahmed2015selectively}.

A boundary element method was employed to examine the pumping and swimming of these three configurations as a function of the semi-coverage of the catalytic end-caps, and different optimal semi-coverages were identified for each configuration. Whilst these quasi-static results represent a numerical proof-of-concept of this hypothesis, it will be important for future studies to develop a full chemo-elastohydrodynamic theory that solves the time-dependent motion of the swimmer, particularly when switching between modes. Such a theory would likely employ a phoretic slender body theory coupled with nonlinear beam mechanics, rather than the boundary element method, to increase numerical efficiency. Once created, the interactions between multiple flexible phoretic filaments can be studied, which will open the doorway to interesting collaborative dynamics, such as the guiding and joining the dynamic self-assembly~\citep{wykes2016dynamic} of rigid particles into transforming superstructures.

%
%
%
%
%

\begin{acknowledgments}
This work is funded by EPSRC grant EP/R041555/1. The author would like to thank Eric Lauga, Sebastien Michelin, and John Meyer for insights and discussion.	
\end{acknowledgments}


%

\end{document}